# All-optical dynamic modulation of spontaneous emission rate in hybrid optomechanical cavity quantum electrodynamics systems

Feng Tian,[1,2,3,*] Hisashi Sumikura,[1,2] Eiichi Kuramochi,[1,2] Masato Takiguchi,[1,2] Masaaki Ono,[1,2] Akihiko Shinya,[1,2] and Masaya Notomi[1,2,3,*]

[1]NTT Basic Research Laboratories, NTT Corporation, 3-1 Morinosato Wakamiya, Atsugi, Kanagawa 243-0198, Japan
[2]Nanophotonics Center, NTT Corporation, 3-1 Morinosato Wakamiya, Atsugi, Kanagawa 243-0198, Japan
[3]Department of Physics, Tokyo Institute of Technology, 2-12-1-H55 Ookayama, Meguro, Tokyo 152-8550, Japan
[*]notomi@phys.titech.ac.jp; Tian_Feng_x7@lab.ntt.co.jp

**Recent nanofabrication technologies have miniaturized optical and mechanical resonators, and have led to a variety of novel optomechanical systems in which optical and mechanical modes are strongly coupled. Here we hybridize an optomechanical resonator with two-level emitters and successfully demonstrate all-optical dynamic control of optical transition in the two-level system by the mechanical oscillation via the cavity quantum-electrodynamics (CQED) effect. Employing copper-doped silicon nanobeam optomechanical resonators, we have observed that the spontaneous emission rate of excitons bound to copper atoms is dynamically modulated by the optically-driven mechanical oscillation within the time scale much shorter than the emission lifetime. The result is explained very well with an analytical model including the dynamic modulation of the Purcell effect and the exciton population. To the best of our knowledge, this is the first demonstration of a dynamic modulation of the spontaneous emission rate by mechanical oscillations. Our achievement will open up a novel field of hybrid optomechanical CQED systems in which three body--optical transitions, optical resonance modes, and mechanical resonance modes--are strongly coupled and will pave the way for novel hybrid quantum systems.**

## Introduction

Cavity optomechanics is an emerging field in photonics, and has been extensively studied in the last decade [1-7]. It is based on micro/nano-mechanical structures coupled to optical cavities, which enable strong coupling between photons and mechanical oscillations. For these studies, it is essential to reduce the feature size of the optomechanical cavities to enlarge the optomechanical interactions. It is also well known that the optical transition of two-level systems (TLS) in miniaturized optical cavities can be tailored by the modified photonic density of states, and this approach is generally referred to as cavity quantum electrodynamics (CQED). If we hybridize the cavity optomechanics with TLS, it will become possible to control the optical transition via the mechanical oscillations as illustrated in Fig. 1(a). Recently, such hybrid optomechanical CQED systems have been discussed theoretically [8-12]. Although the electromechanical tuning of a photonic crystal (PhC) cavity has already been employed to *statically* modify CQED systems [13,14], the recent success of cavity optomechanics suggests that we can control the optical transition *dynamically* within the time scale of the transition, which is required for hybrid optomechanical CQED systems, and this will open up a novel field for optomechanics and CQED.

An important manifestation of this system is the dynamic control of the spontaneous emission (SE) rate based on the Purcell effect [15-17]. Besides optomechanics, the dynamic control of the SE rate via the Purcell effect has been attracting attention and was recently demonstrated with solid-state CQED systems using several methods including the dynamic modification of cavity loss by optical pumping [18], the electrical tuning of exciton energy via the Stark effect [19], and the acousto-optic control of quantum dots (QDs) in PhC cavities [20]. In addition, there have been studies involving the tuning of TLS by strain manipulation [21-25], but the strain can only change the transition frequencies. Without the CQED effect, the transition rate is hardly modified. In fact, a dynamic SE rate modification in a mechanical system has yet to be reported. The purpose of the present work is to realize dynamic control of the SE rate by mechanical oscillation through a strongly coupled optomechanical CQED system.

The principle of optomechanical SE rate modulation is illustrated in Fig. 1(b), where the cavity resonance is optomechanically modulated around the transition frequency of the TLS. When the cavity resonance is on the transition frequency, the SE rate is enhanced by the Purcell effect and thus the intensity becomes stronger. When the cavity resonance is detuned from the transition frequency, the enhancement vanishes. Therefore, if the cavity is dynamically modulated by the mechanical oscillation, the SE rate will be dynamically modulated. As far as we know, this operation has never been achieved, and is our primary target for this study.

To achieve this goal, spectral and temporal requirements have to be met. First, in the optical spectral domain, the modulated cavity resonance wavelength must be able to completely cross the emission line. For this purpose, the emission linewidth should be as narrow as possible and the modulation amplitude of cavity resonance wavelength should be sufficiently large. Second, in the time domain, the dynamic modification of the SE rate should be faster than the intrinsic SE decay. This means that the cavity resonance should cross the emission line at a speed exceeding the SE rate. It is worth noting that the optical transition rates of nano-emitters are normally orders of magnitudes higher than the mechanical resonance frequencies [13], which makes the current target difficult. In addition, if the mechanical resonance frequency is much higher, the displacement amplitude will be much smaller [3], thus making the wavelength tuning range too small to cover the emission line.

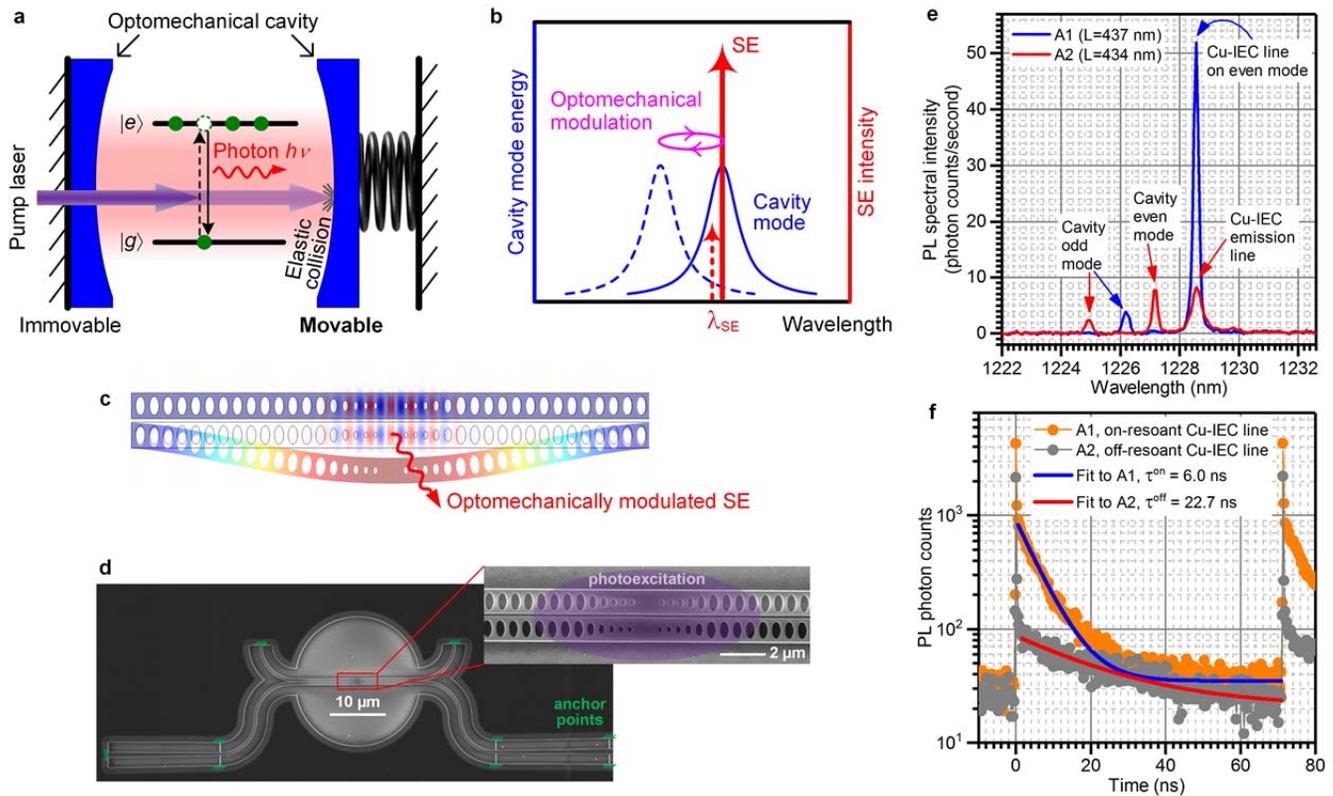

**Figure 1 | Hybrid optomechanical cavity quantum electrodynamics (CQED) system with the dynamic Purcell effect. a,** Schematic of optomechanized CQED system, where the optical cavity is movable and correspondingly the cavity resonance is tunable. A pump laser is used to excite photoluminescence (PL) of two-level emitters inside the cavity and simultaneously drive the movable structure of the system. **b,** Spectral illustration for the system in **a**, where the cavity resonance is optomechanically modulated and in turn dynamically enhances the spontaneous emission (SE) via the Purcell effect. **c,** Concrete scheme of optomechanical configuration showing the finite-element-method (FEM) simulated mechanical (fundamental in-plane) and optical (fundamental even) resonance modes. **d,** Scanning electron microscope (SEM) image of a fabricated silicon (Si) optomechanical cavity device. Copper isoelectronic centers (Cu-IECs) have been formed as the two-level emitters by doping the Si layer with copper atoms before we fabricate the nano-structures. Inset: illustration of the focused pump laser (375 nm) for photoexcitation of the Cu-IEC emitters in cavity region. **e,f,** Experimental verification of the Purcell effect for this cavity design with immovable devices A1 and A2, which have slightly different cavity lengths (L). Their PL spectra are shown in **f**, and the corresponding time-resolved PL decay curves of the zero-phonon emission line ($\lambda_{\text{Cu-IEC}}$) are shown in **g**.

To satisfy all the above criteria, we employ copper dopants in silicon (Si) as a TLS [26-28], which constitutes a key issue for the present study. Copper dopants in Si are known to form isoelectronic centers (IECs), which are extraordinarily-efficient emitters in Si [29] and exhibit very sharp emission line (a linewidth of ~0.2 nm or less even for ensemble of Cu-IECs in Si [26-28]). They are thus considered an ideal solid-state TLS with which to observe the CQED effect in Si. Previously we demonstrated accelerated radiative decay caused by the Purcell effect for Cu-IECs in Si PhC nanocavities [28]. The intrinsic photoluminescence (PL) decay time of Cu-IECs in Si is ~30 ns, which is much longer than those of various quantum dots (QDs) in III/V semiconductors (0.1-1 ns) [13,16], thereby making the present optomechanical modulation experiment much easier to perform at moderate mechanical frequencies. In addition, the dopants in Si generally have some potential to be employed as qubits in quantum computing [30] where the CQED devices will play key roles [31]. The use of Si is also advantageous because of its mature nanofabrication technologies, which allow us to realize a high quality factor ($Q$) optomechanical cavity, which is also very important for the present work.

Another key issue is the optomechanical cavity in which the Cu-IEC emitters are embedded. Here, we adopt a doubly coupled Si nanobeam PhC cavity with the mechanical and optical resonance modes illustrated in Fig. 1(c). Generally, double-beam (or double-layer) PhC nanocavities exhibit exceptionally large optomechanical coupling strength because of their large optical field gradient in the gap between two cavities [32-37], and here we specially design the cavity for a CQED system without any deterioration in the optomechanical coupling (see Supplementary Fig. s1).

In this paper, we report our successful demonstration of the dynamic modulation of the SE intensity and rate for Cu-IECs in Si optomechanical CQED systems. We drive mechanical oscillation of the nanobeam by using a pulsed laser [38,39], which is simultaneously used to excite the PL of the emitters. Thus, our experiment is operated all-optically. We believe this to be the first experimental demonstration of the dynamic control of SE rate by using cavity optomechanics via the Purcell effect. In other words, we have realized dynamically-controlled hybrid optomechanical CQED systems for the first time. We believe our result will pave the way to novel fields in optomechanics based on hybrid optomechanical CQED systems, such as novel phenomena of tripartite interactions between quantum emitter, optical cavity and mechanical oscillator [12].

## Results

### 1. Purcell effect in immovable cavity

Our optomechanical cavity systems are based on the doubly coupled nanobeam PhC cavity shown in Fig. 1(c). Each beam has a defect consisting of missing holes in the central PhC region, which is specially designed to keep a sufficiently large space for emitters. The optical design is detailed in Supplementary Fig. s1. The devices are fabricated in two steps. First, the Cu-IECs are uniformly doped into the device layer of a silicon-on-insulator (SOI) wafer by ion implantation and rapid thermal annealing [28]. Second, the structures are fabricated with a conventional nanofabrication process (see details in Method). A scanning electron microscope (SEM) image of the device is shown in Fig. 1(d).

As a prerequisite, we first verify the Purcell effect for Cu-IEC emitters in our double-beam cavities by employing the same optical design but mechanically immovable cavities, which are mechanically anchored between two coupled nanobeams (see Supplementary Fig. s3). Figure 1(e) shows the PL spectra of two immovable cavities (A1 and A2) with slightly different cavity lengths, where the cavities' odd and even resonance modes are identified according to finite element method (FEM) simulations. Note that the even mode of A1 overlaps the zero-phonon emission line of the Cu-IECs (Cu-IEC line for abbreviation) at 1228.6 nm ($\lambda_{\text{Cu-IEC}}$). The fine wavelength alignment method is described in Supplementary Fig. s5. We are interested in this even mode because it has much stronger optomechanical coupling than the odd mode [36,37]. The PL intensity of A1 is enhanced ~6.5 times compared with that of A2 in which the Cu-IEC line is out of cavity resonance. The PL lifetime obtained by time-resolved PL decay measurements is 6.0 ns for A1, which is greatly accelerated from 22.7 ns for A2 (see Fig. 1(f)). From these two observations, we confirmed that our doubly coupled nanobeam cavity exhibits a strong Purcell effect for Cu-IEC emitters. The present results are essentially similar to those of our previous work on Cu-doped Si two-dimensional PhC nanocavities [28], where we confirmed that the Purcell effect explains the observed PL intensity enhancement.

## 2. Optomechanical characterization of movable cavity

Here, we switch to the movable cavity device (B1) without anchors between nanobeams, and in this section we investigate its optomechanical characteristics. We choose the in-plane fundamental mechanical resonance mode in a longer beam of the coupled cavity (see inset in Fig. 2(a)), whose resonance frequency is designed to be 4.1 MHz. In the present experiment, a pulsed pump laser ($\lambda$ = 375 nm, the pulse duration is 90 ps) is used to excite PL and simultaneously drive the mechanical oscillator (see Fig. 2(a)). The pump laser exerts a periodic force on the mechanical oscillator, and we can excite the mechanical resonance modes by tuning the pulse repetition rate (drive frequency for optomechanical modulation, $f_{\text{mod}}$). As explained in section 3 in Supplementary Information, we confirmed that a repetition rate of around 4 MHz does not cause any thermal accumulation effect under our experimental condition (9.8 pJ/pulse). Hereafter, we employ this pulse energy for pumping.

Figure 2(b) shows a PL spectrum of device B1 excited with $f_{\text{mod}}$ of 4.0 MHz. Since $f_{\text{mod}}$ is off-resonant with the mechanical mode, this result represents an unmodulated cavity resonance wavelength (referred as $\lambda_{\text{cav0}}$). The cavity resonance mode (37.5 GHz wide) is detuned from the Cu-IEC line (58.8 GHz wide) by 1.27 nm (255 GHz). Since the separation is much wider than the width of the two peaks, we regard the Cu-IEC line as being completely decoupled from the cavity resonance under this condition. Next, we examine whether we can dynamically induce resonant coupling by mechanical modulation. Before proceeding to the dynamic modification of SE, we examine the mechanical resonance properties of the device B1 by employing optical transmission measurements. With the pulsed pump laser irradiating the cavity, an auxiliary tunable continuous-wave (CW) probe laser is simultaneously launched into the waveguide, and the transmitted intensity of the probe is measured via an objective lens above the cavity (See Fig. 2(a)). By setting the wavelength of the probe laser ($\lambda_{\text{probe}}$) to $\lambda_{\text{cav0}}$ (1227.15 nm) and sweeping $f_{\text{mod}}$ of the pump laser, we observe radio-frequency (RF) resonance in the transmitted intensity of the probe at around 4.1 MHz as shown by the gray line in Fig. 2(c). Note that the resonance frequency agrees with the theoretical mechanical resonance mode of interest.

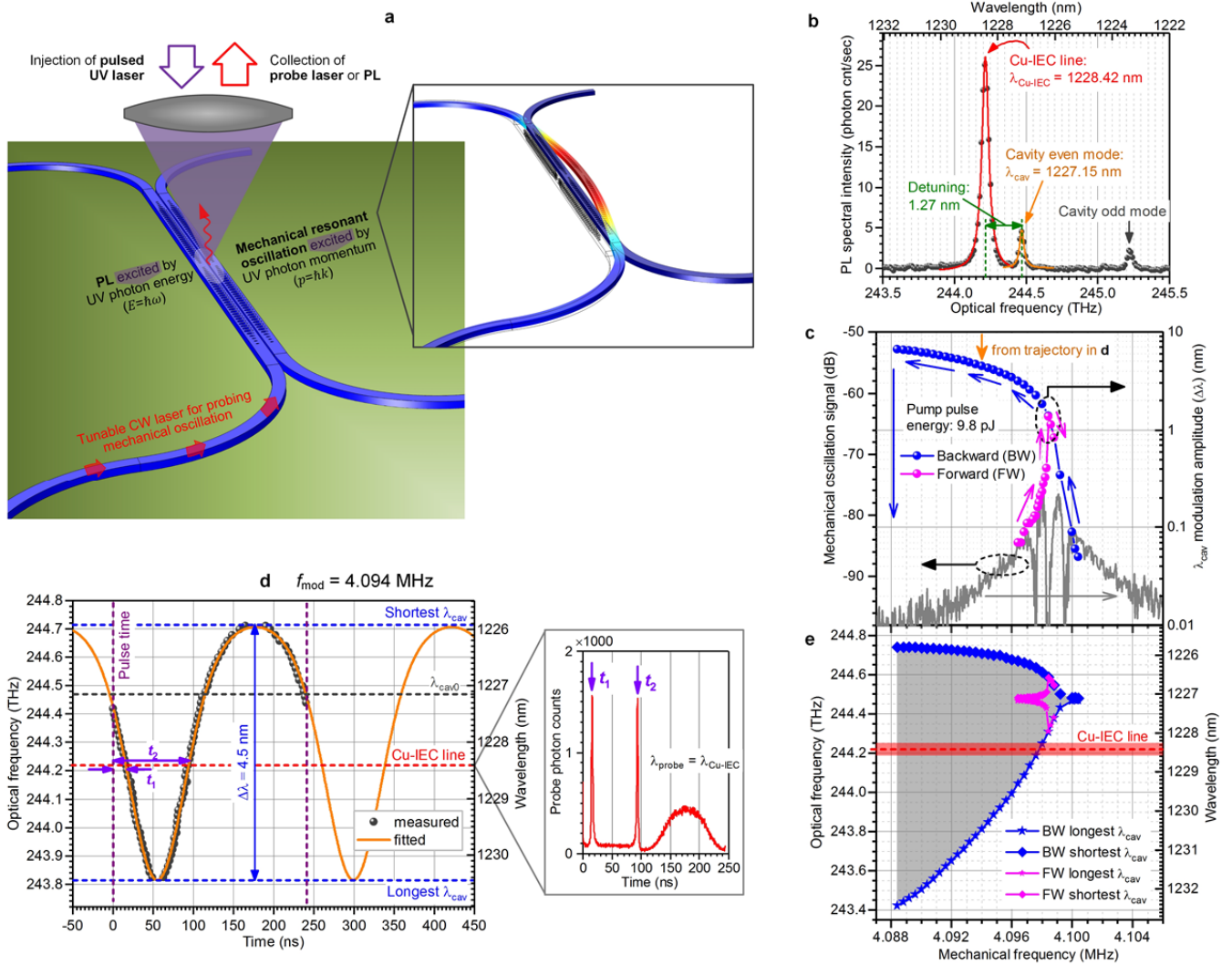

**Figure 2 | Optomechanical characteristics of the device B1 prepared for SE rate modulation. a,** Experimental scheme to excite both the PL and the mechanical resonant oscillation with a single pulsed laser. A probe laser (CW, tunable wavelength ($\lambda_{probe}$)) is launched into the cavity along the waveguide to assist characterization. Inset: FEM simulated mechanical resonance mode of one nanobeam of the B1. **b,** PL spectrum of the B1 excited with a pulse repetition rate ($f_{mod}$) of 4 MHz and a pulse energy of 9.8 pJ/pulse. **c,** Modulation amplitude ($\Delta\lambda$) of cavity resonance (even mode) wavelength ($\lambda_{cav}$) as a function of $f_{mod}$, which shows a strong Duffing softening nonlinearity. The Gray curve is the oscillation signal directly measured by the probe laser (note that we set $\lambda_{probe}$ equals unmodulated cavity resonance wavelength ($\lambda_{cav0}$)). **d,** Evolution of $\lambda_{cav}$ in one mechanical oscillation period at $f_{mod}$ = 4.094 MHz. Measured data are fitted by a combination of the simulated optomechanical coupling coefficient ($g_{OM}$) and a simple harmonic motion. Inset: temporal probe signal acquired by superconducting single photon detector (SSPD), where the times ($t_1$ and $t_2$) that $\lambda_{cav}$ crosses $\lambda_{probe}$ (note that we set $\lambda_{probe}$ equals $\lambda_{Cu-IEC}$ here) are determined. **e,** Modulation range (gray area) of $\lambda_{cav}$ as a function of $f_{mod}$. Red area represents width of the Cu-IEC line.

Next, we measure the time-resolved transmitted intensity. The inset in Fig. 2(d) shows the temporal transmitted intensity at $f_{mod} = 4.094$ MHz and $\lambda_{probe} = \lambda_{cav0}$. We can see that the signal twice increases sharply at $t_1$ and $t_2$. As shown in Supplementary Fig. s11, the appearance of these two peaks continuously changes as $\lambda_{probe}$ is changed. The two peaks indicate that the modulated cavity resonance wavelength ($\lambda_{cav}$) crosses $\lambda_{probe}$ twice during one mechanical oscillation cycle. If this is the case, the timing of the peaks should change in a sinusoidal-like manner with $\lambda_{probe}$ because of the simple harmonic motion of the nanobeam oscillation. This is exactly what we see in Fig. 2(d) in which we plot $t_1$ and $t_2$ (black dots) extracted from the temporal data at different $\lambda_{probe}$ values. This result indicates that we successfully excited the targeted mechanical resonant oscillation by optical driving.

The curve in Fig. 2(d) is not purely sinusoidal. This non-sinusoidal behavior can be expected because $\lambda_{cav}$ is not rigorously linear in relation to the oscillator's displacement [36,37]. This relation is represented by the optomechanical coupling coefficient $g_{OM}/2\pi = d\nu_{cav}/dx_{slot}$, where $\nu_{cav} = c/\lambda_{cav}$ ($c$ is the speed of light in vacuum) is the cavity resonance frequency and $x_{slot}$ is the slot gap between nanobeams [33]. $g_{OM}/2\pi$ as a function of $x_{slot}$ is estimated by FEM simulations to be around 10 to 40 GHz/nm (see Supplementary Fig. s11), and thus the $\nu_{cav}$ as a function of $x_{slot}$ is derived by $\nu_{cav}(x_{slot}) = \int \frac{g_{OM}(x_{slot})}{2\pi} dx_{slot}$. By combining $\nu_{cav}(x_{slot})$ and a simple harmonic motion model, the observed trajectory is accurately fitted as shown in Fig. 2(d). This proves that this trajectory represents the temporal variation of $\lambda_{cav}$ induced by the targeted optomechanical resonant oscillation. It may be worth noting that $\lambda_{Cu\text{-}IEC}$ is located within the modulation range of $\lambda_{cav}$, and the corresponding $t_1$ ($\approx 16$ ns) is shorter than the PL decay time, so the device B1 is promising for use in SE rate modulation experiments.

From the trajectory curve in Fig. 2(d), we can extract the $\lambda_{cav}$ modulation amplitude ($\Delta\lambda$) which is 4.5 nm. In Fig. 2(c), we plot $\Delta\lambda$ extracted from a series of measurements similar to Fig. 2(d) with continuously varying $f_{mod}$. This amplitude should correspond to the temporal variation of $\lambda_{cav}$ induced by mechanical oscillation, and therefore should be approximately proportional to the displacement amplitude of the oscillation. As shown in the figure, the observed amplitude shows peculiar hysteresis as a function of $f_{mod}$, which is typical behavior of the displacement amplitude for a Duffing nonlinear oscillator [40,41]. As a result of the hysteresis of Duffing softening nonlinearity, $\Delta\lambda$ is large when $f_{mod}$ is backward swept as seen in Fig. 2(c). In fact, the trajectory in Fig. 2(d) was obtained under this condition (as noted by the orange arrow in Fig. 2(c)). This Duffing nonlinearity gives us an extra degree of freedom in the SE modulation experiments. We can increase the displacement amplitude by choosing an appropriate branch in the hysteresis as shown above. Figure 2(e) shows the modulation range of $\lambda_{cav}$ within one cycle (longest and shortest $\lambda_{cav}$) as a function of $f_{mod}$. Note that $\lambda_{Cu\text{-}IEC}$ falls into the modulation range when $f_{mod}$ is between 4.098 and 4.088 MHz in a backward sweep.

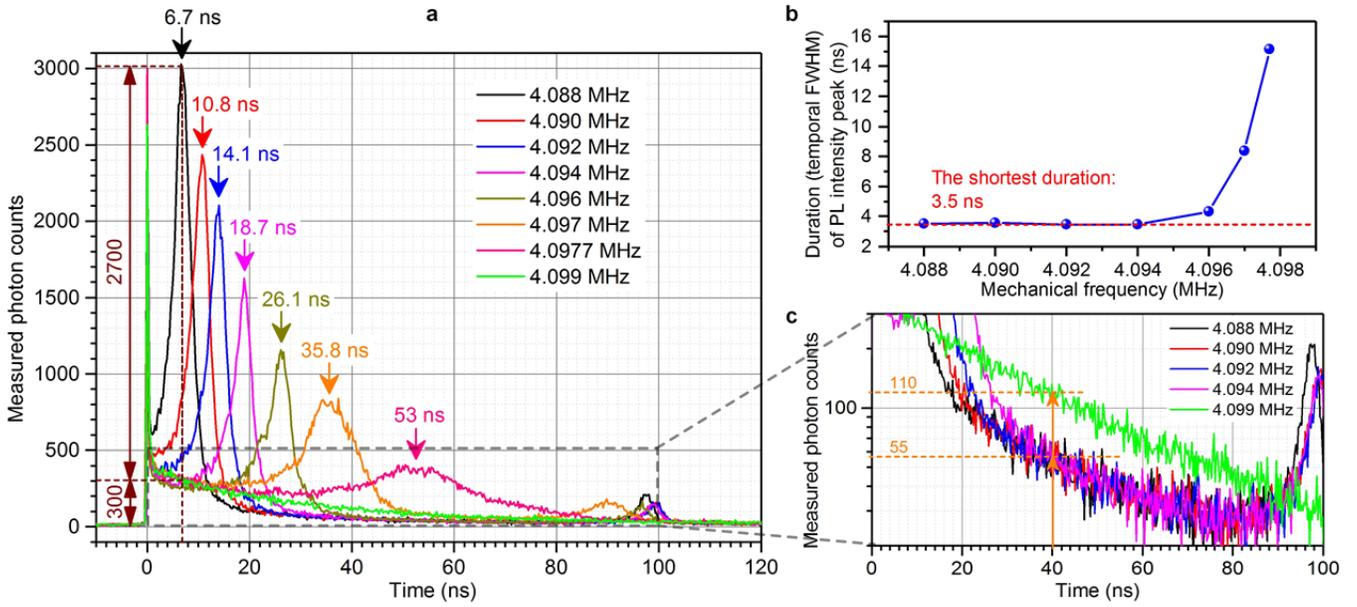

**Figure 3 | Optomechanically modulated PL decay curves at various $f_{mod}$ values. a,** PL intensities are enhanced when the $\lambda_{cav}$ crosses $\lambda_{Cu\text{-}IEC}$. Analyses indicate that these enhancements can be mainly attributed to the Purcell effect. **b,** Duration (temporal FWHM) of the PL intensity peak at $t_1$ as a function of $f_{mod}$. **c,** After $\lambda_{cav}$ crosses $\lambda_{Cu\text{-}IEC}$, fewer photons are emitted in decay curves at frequencies of 4.088 to 4.094 MHz than in the decay curve at 4.099 MHz where $\lambda_{cav}$ has never reached to $\lambda_{Cu\text{-}IEC}$. This is because more excitons were consumed at the Purcell enhanced radiative recombination rate.

## 3. Dynamic SE control in hybrid cavity QED system

[*Main experimental results*]

Here we remove the probe laser and measure the emission dynamics of our optomechanical CQED system, which is our final target. Figure 3(a) shows the results of time-resolved PL decay measurements for device B1 at different $f_{mod}$ values with a backward sweep (the pump laser condition is the same as that in Fig. 2). When $f_{mod} = 4.099$ MHz, at which the modulation amplitude is very small and $\lambda_{cav}$ cannot reach to $\lambda_{Cu\text{-}IEC}$ at all, we observed the conventional monotonic exponential decay of the PL intensity, similar to the off-resonant PL decay of the immovable cavity (A2) shown in Fig. 1(f). However, when $f_{mod}$ is varied from 4.0977 to 4.088 MHz, the conventional PL decay curve is interrupted by a sudden sharp increase in the PL intensity at a certain elapsed time after the pulse excitation. For example, when $f_{mod} = 4.094$ MHz , there are two peaks along the decay curve and their elapsed times (18.7 and 100 ns) agree well with the $t_1$ and $t_2$ at $\lambda_{cav} = \lambda_{Cu\text{-}IEC}$ in Fig. 2(d), respectively, which strongly suggests that these increases in temporal emission during the PL decay measurements originate from the dynamic coupling of the cavity resonance to the Cu-IEC emission line.

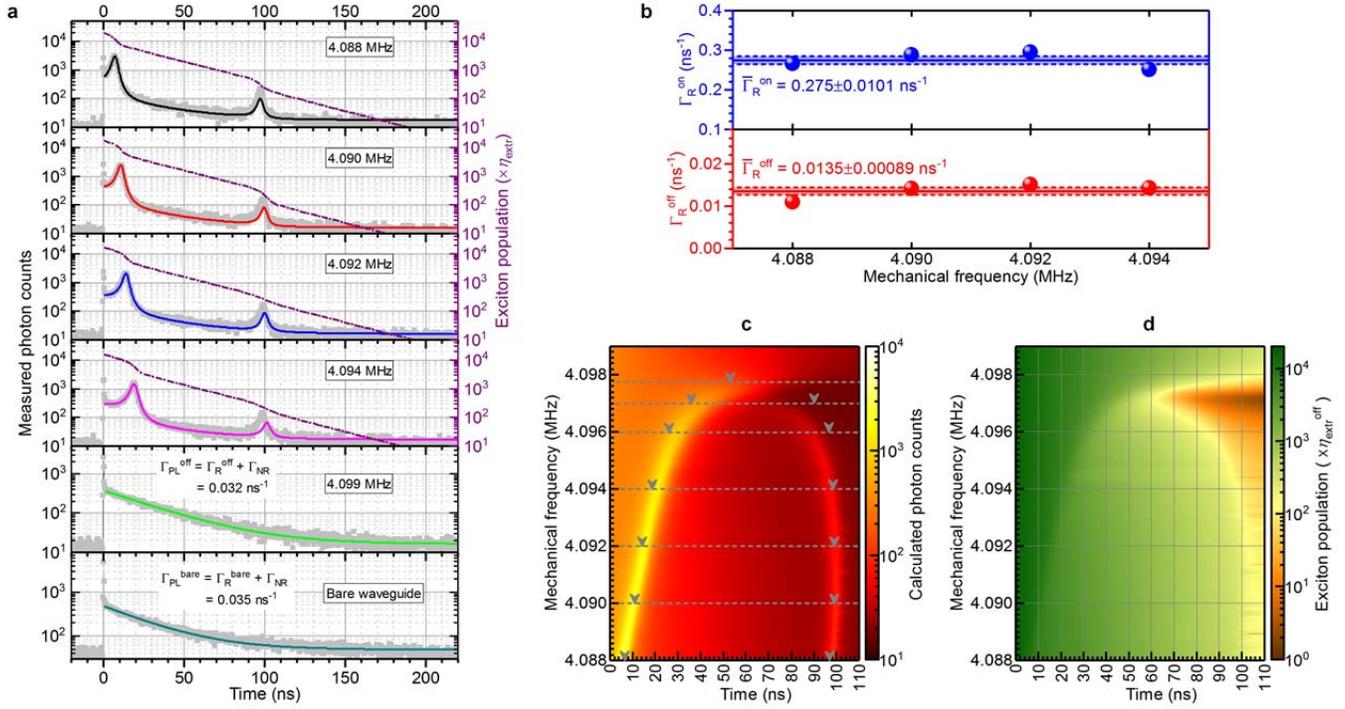

**Figure 4 | Analytical model for dynamics of PL intensity and exciton population in our hybrid optomechanical CQED system. a,** Fittings for the measured PL decay curves in Fig. 3a with the analytical model (see details in section 5.2 in Supplementary Information) and the correspondingly calculated exciton population dynamics. Off-resonant PL decay curve of the B1 cavity at $f_{\text{mod}}$ = 4.099 MHz and PL decay curve of a neighboring bare waveguide ($f_{\text{mod}}$ = 4.099 MHz) are utilized to deduce experimental Purcell factor which is 16.5. **b,** Radiative decay rates on (upper) and off (lower) the cavity resonance. Scattered data are extracted from the fitting curves in **a**, and horizontal lines with error bars (standard deviation) are the mean values of the scattered data, respectively. **c,d,** Calculated dynamics of PL intensity (**c**) and exciton population (**d**) as functions of $f_{\text{mod}}$ (see calculation parameters in Supplementary Fig. s15), respectively. The measured $t_1$ and $t_2$ in Fig. 3a are marked in **c** for comparison.

Here, we examine the abnormal PL decay curves shown in Fig. 3(a) in more detail in relation to the optomechanical dynamics observed in Fig. 2. First, for $f_{\text{mod}}$ < 4.0977 MHz, there are always two peaks in Fig. 3(a), which can be understood as meaning that $\lambda_{\text{cav}}$ crosses $\lambda_{\text{Cu-IEC}}$ twice in one modulation period for $f_{\text{mod}}$ < ~4.0977 MHz in Fig. 2(e). Second, temporal separation between the two peaks becomes smaller when $f_{\text{mod}}$ approaches 4.0977 MHz. This is consistent with Fig. 2(e) where the modulation amplitude is reduced as $f_{\text{mod}}$ becomes higher. It can be understood from the $\lambda_{\text{cav}}$ trajectory in Fig. 2(d) that the smaller $\Delta\lambda$ induces the shorter $|t_2 - t_1|$ between the two points at $\lambda_{\text{cav}} = \lambda_{\text{Cu-IEC}}$. We extract the times of peaks from Fig. 3(a) and summarize them as a function of $f_{\text{mod}}$ as gray marks in Fig. 4(c). Their behavior in relation to $f_{\text{mod}}$ is very similar to the $\lambda_{\text{cav}}$ modulation amplitude versus $f_{\text{mod}}$ in Fig. 2(e). Consequently, the dynamic PL intensity enhancement in Fig. 3(a) is consistently explained by the optomechanical oscillation in Fig. 2.

It is worth noting that the peak duration at $t_1$ (that is, the temporal full width at half maximum (FWHM) of the peaks in Fig. 3(a), summarized in Fig. 3(b)) becomes shorter as $f_{mod}$ decreases. This tendency can be understood if we look at the evolution of the modulated $\lambda_{cav}$ shown in Fig. 2(d). The peak duration is determined by how fast $\lambda_{cav}$ crosses $\lambda_{Cu-IEC}$, namely $d\lambda_{cav}(t)/dt$ (slope of the curve in Fig. 2(d)) at $\lambda_{cav} = \lambda_{Cu-IEC}$. For the higher $f_{mod}$ in Fig. 3(b), the $\lambda_{cav}$ modulation amplitude is smaller (see Fig. 2(e)), and $\lambda_{Cu-IEC}$ is closer to the minimum of the sinusoidal-like curve where $d\lambda_{cav}(t)/dt$ is zero (see lower half panel of Fig. 2(d)), so the PL intensity enhancement lasts longer. When $f_{mod}$ is changed to the lower side, $\lambda_{Cu-IEC}$ approaches the midpoint of the $\lambda_{cav}$ modulation range, where $d\lambda_{cav}/dt$ is at its largest and relatively linear, so the duration in Fig. 3(b) converges to its shortest value of 3.5 ns. This value is much shorter than the intrinsic PL lifetime of Cu-IECs in Si (~30 ns) and comparable to those of some of III-V semiconductor QDs [13, 14, 22]. This is one of the interesting features of the dynamic control of SE by which we can change the SE properties much faster than its intrinsic radiative recombination rate.

[*Contribution of Purcell effect*]

We have reported that the PL intensity is dynamically enhanced by the optomechanically modulated cavity resonance in Fig. 3(a), but this PL intensity enhancement could be explained by the Purcell effect (SE rate enhancement) and/or the resonance-enhanced photon extraction efficiency [42]. The theoretical Purcell factor of the device B1 is approximately the same as those of A1 and A2 (see Supplementary Table s1), for which we have experimentally verified the unambiguous Purcell effect (Fig. 1(f)). We believe that B1 would exhibit an accelerated PL decay curve similar to that observed in A1 in Fig. 1(g) if we were able to statically tune $\lambda_{cav}$ so that it overlapped $\lambda_{Cu-IEC}$, but we have no experimental means to do so.

In addition to the above speculation, we investigate Fig. 3(a) more carefully to distinguish the Purcell effect and the extraction enhancement. Figure 3(c) is a magnified plot of Fig. 3(a). We note that the PL intensities between $t_1$ and $t_2$ at $f_{mod} = 4.088$ to $4.094$ MHz are obviously smaller than the intensity at the same time at $f_{mod} = 4.099$ MHz where we observe no enhancement. This reduction can only be explained by the Purcell effect. For the same initial exciton population, the Purcell enhancement should lead to a significant reduction in the exciton population after the first PL peak at $t_1$ because of the enhanced radiative recombination rate. In contrast, extraction efficiency enhancement does not alter the exciton population dynamics, and thus the PL intensity after the first peak should be the same as that at $f_{mod} = 4.099$ MHz. On the other hand, from an FDTD simulation, we numerically estimate extraction efficiency as a function of wavelength detuning (see Supplementary Fig. s14). The ratio of the extraction efficiencies on/off the cavity resonance is 1.8:1, which is significantly smaller than the enhancement ratio (9:1) of PL intensity observed in Fig. 3(a). Thus, the Purcell effect dominantly plays a role in dynamically enhancing the PL decay curves.

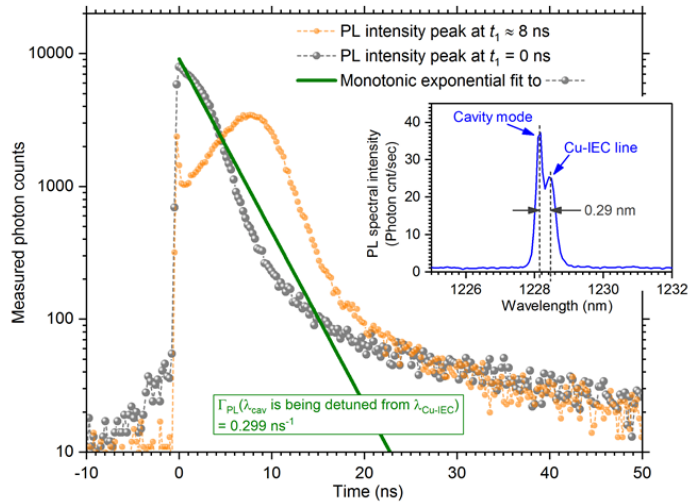

**Figure 5 | PL decay curve with dynamically-detuned Purcell effect in case of $t_1$=0.** Here, device B2 is measured, which is neighboring B1 on the same sample and is different from it only in cavity length (467 nm for B2 and 458 nm for B1). Inset: PL spectrum of the B2 showing a much smaller initial separation of 0.29 nm between $\lambda_{cav0}$ and $\lambda_{Cu-IEC}$. Orange dot curve: modulated PL decay curve at $f_{mod}$ = 4.0975 MHz. As observed, the PL intensity peak is at $t_1$ of ~8 ns. Gray dot curve: modulated PL decay curve when $t_1$ equals zero by adjusting $f_{mod}$ to 4.0966 MHz. Green solid curve: monotonic exponential fitting for the gray dots to derive an estimated PL decay rate of 0.299 ns⁻¹.

[*Analytical model for system*]

To investigate the dynamic Purcell effect quantitatively, we employ an analytical model for the PL dynamics in our hybrid optomechanical CQED system. We solve time-dependent equations for the exciton population and emission intensity, where Purcell enhancement is optomechanically modulated by the mechanical oscillation at $f_{mod}$. The modeling is described in detail in section 5.2 in Supplementary Information. We determined that the measured time-resolved PL decay curves in Fig. 3(a) can be fitted very well by this analytical model as shown in Fig. 4(a). The exciton population dynamics are also shown in the same figure, which clearly shows that the population decays become remarkably faster at the moments of the PL peaks. From these fittings, we can estimate the on-resonant and off-resonant radiative recombination rates ($\Gamma_R^{on}$ and $\Gamma_R^{off}$), summarized in Fig. 4(b). Their mean values are $\overline{\Gamma_R^{on}}$ = 0.275 ns⁻¹ and $\overline{\Gamma_R^{off}}$ = 0.014 ns⁻¹, respectively. By using $\Gamma_{PL}^{off} = \Gamma_R^{off} + \Gamma_{NR}$ = 0.032 ns⁻¹, which is easily deduced from the PL decay curve at $f_{mod}$ = 4.099 MHz, we derive $\Gamma_{NR}$ = 0.019 ns⁻¹. Similarly, by using $\Gamma_{PL}^{bare} = \Gamma_R^{bare} + \Gamma_{NR}$ = 0.035 ns⁻¹, which is obtained from the reference bare waveguide measurement, $\Gamma_R^{bare} \approx$ 0.017 ns⁻¹. Combining them all, we determine that the experimental Purcell factor ($F_P = \Gamma_R^{on}/\Gamma_R^{bare}$) is 16.5 for the present optomechanical CQED system.

We evaluate the dynamic Purcell enhancement as a function of $f_{mod}$ using the analytical model. The calculation results are shown as continuously-varying PL peaks (see Fig. 4(c)) and exciton population drops (see Fig. 4(d)) with $f_{mod}$, respectively. The measured discrete $t_1$ and $t_2$ (gray marks in Fig. 4(c)) extracted from Fig. 3(a) agree with the calculated values in Fig. 4(c), which demonstrates the reliability of this model for our hybrid optomechanical CQED system. As shown in Fig. 4(c,d), the timing of the dynamic Purcell effect is continuously adjustable by $f_{mod}$. This feature benefits from the nonlinearity of the mechanical oscillator and is an advantage of our system.

[$t_1$=0 case: *dynamically-detuned Purcell effect*]

Thanks to the very fast optomechanical modulation, the PL peak duration of 3.5 ns in Figs. 3(a) and (b) is shorter than the on-resonant PL lifetime of 6 ns in Fig. 1(f). Thus, if the PL peak is shifted to the time zero (pump pulse time), we can expect a faster decay rate of the PL intensity than the static on-resonant PL decay rate. In this case, the decay process is controlled by a dynamically-detuned Purcell effect where $\lambda_{cav}$ is being detuned from $\lambda_{Cu-IEC}$. We perform this experiment with device B2 for its smaller initial separation of $\lambda_{cav}$ and $\lambda_{Cu-IEC}$ (see inset in Fig. 5) and set its $t_1$=0 ns by adjusting $f_{mod}$. The result is a monotonic decay curve as shown in Fig. 5. It is approximately twice as fast as the statically Purcell-enhanced PL decay curve in Fig. 1(f). This can be explained by the quantum efficiency (see Supplementary Eq. s10) which is drastically decreasing with the detuning of $\lambda_{cav}$ from $\lambda_{Cu-IEC}$. This technology will find application in ultrafast modulation of the on-chip light sources, which is usually limited by the decay rates of the emitters [43].

## Discussion

In this study, we demonstrated the experimental observation of the optomechanical modulation of SE intensity and rate by a mechanical oscillator, and built an analytical model to unveil the underlying dynamics of this hybrid optomechanical CQED system. In the system, three body--two-level emitters, optical cavity and mechanical oscillator--were linked by both the cavity optomechanical and CQED (the Purcell effect) effects. Here, figures of merit (FOMs) of the two effects--$g_{OM}/2\pi$ and $F_P$--are 10 to 40 GHz/nm and 16.5, respectively. This is a good sample for future design of the hybrid optomechanical CQED systems, because the high FOM values ensure a strong coupling of the system. Our experiment also benefited from Cu-IECs in Si which have a long emission lifetime and a narrow linewidth. These properties allow a sufficient modulation of the SE by a mechanical oscillator with moderate frequency but large displacement. Moreover, dopants in Si are promising as qubits for upcoming quantum technologies [31], so we can expect mechanical manipulation of these qubits by the hybrid optomechanical CQED systems. The experimental scheme of pulsed optical force used here is feasible for optomechanical devices from classical to quantum regimes [7,39], therefore, the recent theoretical discoveries in hybrid optomechanical CQED systems [8-12] may find experimental verifications based on our scheme. The hybrid optomechanical QED systems will also inspire some technologies for future industrial applications, for example, the mechanically $\lambda$-tunable nano-lasers and the mechanically switchable on-chip single-photon sources.

## Acknowledgments

This project was supported by the Japan Society for the Promotion of Science (JSPS) (KAKENHI grant No. 15H05735). The authors thank S. Sergent and W. J. Munro for valuable discussions. The authors thank K. Nozaki and G. Zhang for help with measurements. The authors thank O. Moriwaki for help with device fabrication.

## Contributions

M. N. conceived and supervised this project. F. T. conceived the experiments and designed the device. H. S. prepared the Cu-doped SOI wafers. E. K., M. O. and F. T. fabricated the device. F. T., H. S. and M. T. performed the measurements. F. T. analyzed the results. A. S. supported the optical design and measurement. M. N. and F. T. wrote the manuscript. H. S. and E. K. commented the manuscript. All the authors have read and checked the manuscript.

## Competing interests

The authors declare no competing interests.

**Method**

**Sample**. We used the SOI wafer with a 300 nm device layer (intrinsic Si) and a 2 μm buried oxide (BOX) layer. The Cu-IECs were doped into device layer of the whole wafer by two steps. First, [63]Cu ions were uniformly implanted into the Si device layer with an acceleration voltage of 100 kV. The implantation dose was $5 \times 10^{13}$ cm[-2]. The distribution of Cu ions was broad and peaked at a depth of 80 nm. Second, a rapid thermal annealing process was employed to convert the implanted Cu ions into Cu-IECs in Si. The wafer was annealed for 30 second at 700 ℃ and cooled down at a speed of 50 K/s.

**Device fabrication**. First, the patterns of the device structures were written by electron beam lithography (EBL). Second, the nanostructures were etched through the device layer by C4F8/SF6 reactive-ion etching. Third, the wafer was diced into small chips, and then the chips were put into hydrofluoric (HF) vapor machine to release the nanostructures from the BOX layer.

**Measurement.** Three types of measurements were performed in this project.

1. Conventional PL measurement (see Supplementary Fig. s3). A pulsed UV laser (375 nm) was focused on the center of the cavity to excite PL. The PL was collected by the objective lens and then passed through the LPF and the polarizer to eliminate the residual UV and select the TE cavity modes, respectively. Finally, one branch of the PL was launched into a spectrometer, and the other was detected by an SSPD after a BPF (central wavelength: 1228.4 nm, bandwidth: ~1 nm) to record the time-resolved PL decay. No optomechanical effect was involved during this measurement.

2. Optomechanical characterization (see Supplementary Fig. s10). In this experiment, we measured a probe laser ($\lambda_{\text{probe}}$) transmitting through the device rather than the PL. First, the RF signal from the VNA swept $f_{\text{mod}}$ of the pulsed laser to search for the mechanical resonance mode. The optomechanically modulated probe intensity was detected by a PMT and then the AC electrical signal from the PMT was recorded by the VNA to plot a mechanical resonance spectrum. Second, at a certain $f_{\text{mod}}$ around the resonance frequency of the mechanical oscillator, the SSPD recorded the time-resolved probe intensity with $\lambda_{\text{probe}}$ swept around $\lambda_{\text{cav}}$, so that the optomechanical evolution of $\lambda_{\text{cav}}$ in the time domain was calibrated (see Fig. 2d).

3. Measurement of the optomechanically modulated PL decay curves (see Supplementary Fig. s15). In this experiment, the probe laser was removed. The $f_{\text{mod}}$ was swept around the resonance frequency of the mechanical oscillator, and the time-resolved PL decay, which was optomechanically modulated via the dynamic Purcell effect, was recorded by the SSPD.

**Analytical model.** We analyzed the dynamic process of this optomechanical CQED system with an analytical model (Eqs. (1-3)). The PL intensity (or measured photon counts) in Fig. 3(a) can be expressed by

$$N_{\text{ph\_meas}}(t) = -\eta_{\text{extr}}(\Delta\omega)\eta_{\text{qe}}(\Delta\omega)\frac{\partial N_{\text{ex}}(t)}{\partial t},$$ (1)

where $\eta_{\text{extr}}$, $\eta_{\text{qe}}$, $N_{\text{ex}}$ and $\Delta\omega = \omega_{\text{cav}} - \omega_{\text{Cu-IEC}}$ are extraction efficiency, quantum efficiency, Cu-IEC exciton population, and frequency detuning of cavity resonance from Cu-IEC line, respectively. The dynamics of the exciton population can be described by

$$\frac{dN_{\text{ex}}(t)}{dt} = -\Gamma_{\text{PL}}(\Delta\omega)N_{\text{ex}}(t), \tag{2}$$

where $\Gamma_{\text{PL}}$ is PL decay rate and can be expanded as

$$\Gamma_{\text{PL}}(\Delta\omega) = \Gamma_{\text{R}}^{\text{on}}(\Delta\omega) + \Gamma_{\text{R}}^{\text{off}} + \Gamma_{\text{NR}}, \tag{3}$$

where $\Gamma_{\text{R}}^{\text{on}}$ and $\Gamma_{\text{R}}^{\text{off}}$ are on-resonant and off-resonant radiative (R) decay rates, respectively; and $\Gamma_{\text{NR}}$ is non-radiative (NR) decay rate. Here, we incorporated the mechanical oscillation into the PL decay model via time-dependent spectral cavity-emitter coupling: $\Delta\omega(t)$. In fact, $\Delta\omega(t)$ has been experimentally acquired in Fig. 2(d). Using this analytical model, we determined the quantitative contribution of the Purcell effect to the PL peaks in Fig. 4(a,b).